\documentclass[aps,prl,twocolumn,groupedaddress]{revtex4-1}
\usepackage{graphicx,color,graphics}
\usepackage{dcolumn} 
\usepackage{amsmath}
\usepackage{amsfonts}
\usepackage{amssymb}
\usepackage{hyperref}
\usepackage{url}
\usepackage{xcolor}

\begin{document}

\title{Demonstration of a dressed-state phase gate for trapped ions}

\author{T. R. Tan}
\email[Electronic address: ]{tingrei.tan@nist.gov}
\author{J. P. Gaebler}
\author{R. Bowler}
\author{Y. Lin}
\author{J. D. Jost}
\altaffiliation{Current address: $\rm{\acute{E}}$cole Polytechnique F$\rm{\acute{e}}$d$\rm{\acute{e}}$rale de Lausanne, Lausanne, Switzerland}
\author{D. Leibfried}
\author{D. J. Wineland}
\affiliation{National Institute of Standards and Technology, 325 Broadway, Boulder, CO 80305, USA}

\date{\today}

\begin{abstract}
We demonstrate a trapped-ion entangling-gate scheme proposed by Bermudez {\it et~al.} [Phys. Rev. A {\bf 85}, 040302 (2012)]. Simultaneous excitation of a strong carrier and a single-sideband transition enables deterministic creation of entangled states. The method works for magnetic field-insensitive states, is robust against thermal excitations, includes dynamical decoupling from qubit dephasing errors, and provides simplifications in experimental implementation compared to some other entangling gates with trapped ions. We achieve a Bell state fidelity of $0.974(4)$ and identify the main sources of error.
\end{abstract}

\maketitle

Trapped ion-based architectures \cite{Cirac95} are promising candidates for constructing a large-scale quantum information processing (QIP) device \cite{bible,Blatt2008,Blatt2012}. Two hyperfine states of trapped ions (with an energy splitting that is first-order insensitive to changes in the magnetic field) provide qubits with coherence times for superposition states exceeding a few seconds \cite{Langer2005,Olmschenk2007,Lucas2008,Benhelm2008PRA}. However, realization of high-fidelity entangling gates on such qubits has proven challenging and the current achievable fidelities are significantly below that required for practical fault-tolerant quantum computation, where the error per gate operation should be below a threshold of around $10^{-4}$ \cite{Preskill98,Knill2010,Ladd2010}.

A number of entangling gates have been proposed and demonstrated for trapped ions, including the Cirac-Zoller gate \cite{Cirac95,Monroe95,SchmidtKaler2003}, and geometric phase gates \cite{Sorenson99,Sorenson2000,Milburn1999,Solano1999}. These gates utilize the coupling between laser beams and internal states of ions, as well as the Coulomb coupling between ions, to create entanglement. The Cirac-Zoller gate requires that the ions be prepared in a pure motional state, which may be difficult to accomplish with errors low enough to be suitable for large-scale QIP. In contrast, the geometric phase gates are relatively insensitive to initial motional states as long as the ions remain in the Lamb-Dicke regime, where the extent of ion motion is much less than the effective wavelength of the excitation fields. With use of a geometric phase gate an error of $3(2) \times 10^{-2}$ was achieved for producing a Bell state of two hyperfine states of $^9$Be$^+$ ions \cite{Leibfried2003}. With use of a geometric phase gate that operates in a rotated basis of the qubit states, as outlined by M\o{}lmer and S\o{}rensen (MS gate) \cite{Sorenson99,Sorenson2000}, an error of $7(1) \times 10^{-3}$ was measured for a Bell state of ``optical" qubits consisting of the ground $S _{1/2}$ and metastable $D_{5/2}$ levels of $^{40}$Ca$^+$ \cite{Benhelm2008}. Of these two gates, only the MS gate can be performed directly on magnetic field-insensitive qubits. \cite{Langer2005,Lee05}. However, in contrast to the situation described in \cite{Benhelm2008} for metastable optical qubits, performing the MS gate on hyperfine qubits with stimulated-Raman transitions requires the use of non-co-propagating beams, and the gate becomes more sensitive to fluctuations of the phases of the laser beams at the ions' positions. While various techniques exist for suppressing the sensitivity of the gate to slow laser path-length fluctuations \cite{Lee05,Gaebler2012}, they require technically demanding laser beam setups and extra (spin-echo) laser pulses; therefore, errors as low as reported in \cite{Benhelm2008} have not been achieved. Previous efforts have been made to combine high-fidelity optical gates with long-coherence hyperfine states, but the fidelities were not as high \cite{Kirchmair2009}. In this work we demonstrate a new entangling phase gate scheme recently proposed by Bermudez {\it et~al.} \cite{Bermudez2012} to achieve a geometric phase gate in a rotated basis using magnetic field-insensitive states, which exhibits reduced technical overhead and improved fidelity relative to the MS gate for hyperfine qubits demonstrated in previous experiments \cite{Home2011,Gaebler2012,Hanneke2009}. This gate is also suitable to be performed with an all-microwave scheme \cite{Ospelkaus2008,Ospelkaus2011}.

The gate requires a carrier spin-flip excitation $|\downarrow ,n \rangle \leftrightarrow |\uparrow ,n \rangle$, and a single spin-motion sideband spin-flip excitation $|\downarrow ,n \rangle \leftrightarrow |\uparrow ,n + 1\rangle$ or $|\uparrow ,n - 1\rangle$, where $n$ is the harmonic oscillator Fock state level of the frequency-selected normal mode of ion motion in the trap, and $|\downarrow\rangle,|\uparrow\rangle$ represent the two internal (here, hyperfine) qubit states \cite{Bermudez2012}. The key concept of this gate scheme is to use a strong carrier excitation to create dressed states of the qubits and apply a spin-dependent force in the dressed states basis using a single laser sideband. The dressed-state nature of the gate reduces the sensitivity of the qubits to dephasing error due to an effective dynamical decoupling resulting from the strong carrier drive \cite{Viola98,Bermudez2012}. Furthermore, the gate scheme has a reduced technical overhead compared to the MS gate due to the use of a single sideband, and is intrinsically insensitive to slow optical path length fluctuation \cite{Lee05}. The carrier excitation, with Rabi frequency $\Omega_{C}$, is set to be resonant with the qubit energy splitting $\omega_0$. A pair of laser beams in a Raman configuration is used to drive a sideband transition detuned by $\delta$ from resonance. Here, the frequency beat note of the beams is set to be $\omega_0 + \omega_{\nu} + \delta$, where $\omega_{\nu}$ is the relevant motional mode frequency. Under the rotating-wave approximation, and in the interaction frame for both the spin and motion, the Hamiltonian for two ions driven by a carrier and a $|\downarrow ,n \rangle \leftrightarrow |\uparrow ,n + 1\rangle$  blue sideband on a single mode of motion can, in the Lamb-Dicke limit, be written as
\begin{eqnarray}
H = \hbar\displaystyle\sum_{j=1,2} \left(\Omega_{C}\sigma_j^+ e^{i\phi} + i \Omega_j \sigma_j^+ a^+ e^{-i\delta t} e^{i\phi^{\prime}_{j}}\right)+h.c.,
\label{eqnZbasis}
\end{eqnarray}
with $\Omega_j = \Omega_0 \eta \xi_j$ where $\Omega_0$ is the Rabi frequency for resonant carrier excitation, $\xi_j$ is the normal mode amplitude of the {\it j}th ion, $\sigma_j^+$ is the spin raising operator, $a$ is the usual ladder lowering operator for the relevant (harmonic) vibrational mode, and $\phi$ and $\phi^{\prime}_{j}$ are the respective phases of the carrier and sideband excitation. The Lamb-Dicke parameter, $\eta$, is equal to $\Delta k_z z_0$, where $z_0=\sqrt{\hbar/2 m \omega_{\nu}}$, $m$ is the mass of a single Beryllium ion and $\omega_{\nu}$ is the normal mode frequency. The distance between the ions is adjusted, by adjusting the strength of the harmonic confinement, to be $\pi p/\Delta k_z$ where $p$ is an integer and $\Delta k_z$ is the difference wave-vector of the Raman laser beams along the axis of motion \cite{Blatt2008}. The carrier excitation can be driven by either a stimulated-Raman process or a microwave field. We will present results for both cases.


Since the interesting case will be when $\Omega_{C}\gg|\Omega_j|$, we go to the interaction frame, where the states are dressed by the carrier excitation, and consider the effects of the sideband terms. For simplicity, we set $\phi=\phi^\prime_j=0$ since they are not crucial for the description below. In the $|+\rangle,|-\rangle$ basis, with $\left|\uparrow\right> = \frac{1}{\sqrt{2}}\left(\left|+\right>+\left|-\right>\right)$ and $\left|\downarrow\right> = \frac{1}{\sqrt{2}}\left(\left|+\right>-\left|-\right>\right)$, the Hamiltonian (\ref{eqnZbasis}) becomes \cite{Bermudez2012}
\begin{eqnarray}
\nonumber H &=& i\hbar\displaystyle\sum_{j} \frac{\Omega_j}{2}\left(\left|+\right>_j\left<+\right|_j - \left|-\right>_j\left<-\right|_j\right)\left(a^+ e^{-i\delta t} - a e^{i\delta t} \right)\\
\nonumber &&+ i\hbar\displaystyle\sum_{j} \frac{\Omega_j}{2}\left(\left|-\right>_j\left<+\right|_j e^{-2 i \Omega_{C} t} - \left|+\right>_j\left<-\right|_j e^{2 i \Omega_{C} t}\right)\\
&&\times\left(a^+ e^{-i\delta t} + a e^{i\delta t} \right).
\label{dressedStatePic}
\end{eqnarray}
%
%
%
The first term of the expression above is a spin-dependent force in the dressed-state basis. The second term induces off-resonant transitions between the dressed-states $\left|+\right>$ and $\left|-\right>$. For $\Omega_{C} \gg \delta$, this term can be neglected in the rotating-wave approximation, and the spin-dependent force can be used to perform a geometric phase gate to generate maximally entangling states \cite{Leibfried2003,Sorenson99,Sorenson2000,Lee05}. The entangled states produced by this gate are insensitive to optical path length changes of the non-co-propagating laser beams occurring on a time scale that is long compared to the gate duration.

A detailed description of the experimental apparatus used can be found in \cite{JostThesis}. Two $^9$Be$^+$ ions are confined along the axis of a linear Paul trap (whose potential is harmonic to good approximation \cite{Home2011}) and have an axial center-of-mass (COM) mode frequency of 2.6 MHz and stretch mode (where two ions oscillate out of phase) frequency of 4.5 MHz. For a single $^9$Be$^+$ ion, the radial secular frequencies are set to be 12.5 and 11.8 MHz. A magnetic field of B = 11.964 mT is applied at $45^\circ$ with respect to the trap axis; at this field, the 2s $^2S_{\frac{1}{2}}$ hyperfine qubit states $\left|F=2,m_F=1\right> = \left|\downarrow\right>$ and $\left|1,0\right> = \left|\uparrow\right>$ states have a splitting, $\omega_0/2\pi = 1.207$ GHz, which is first-order insensitive to changes in the applied magnetic field (a detailed energy level diagram of $^9$Be$^+$ ion can be found in \cite{JostThesis}). Coherence times on the order of several seconds have been observed for superpositions of these states \cite{Langer2005,Gaebler2012archive}.

\begin{figure}
\includegraphics[width=0.6\linewidth]{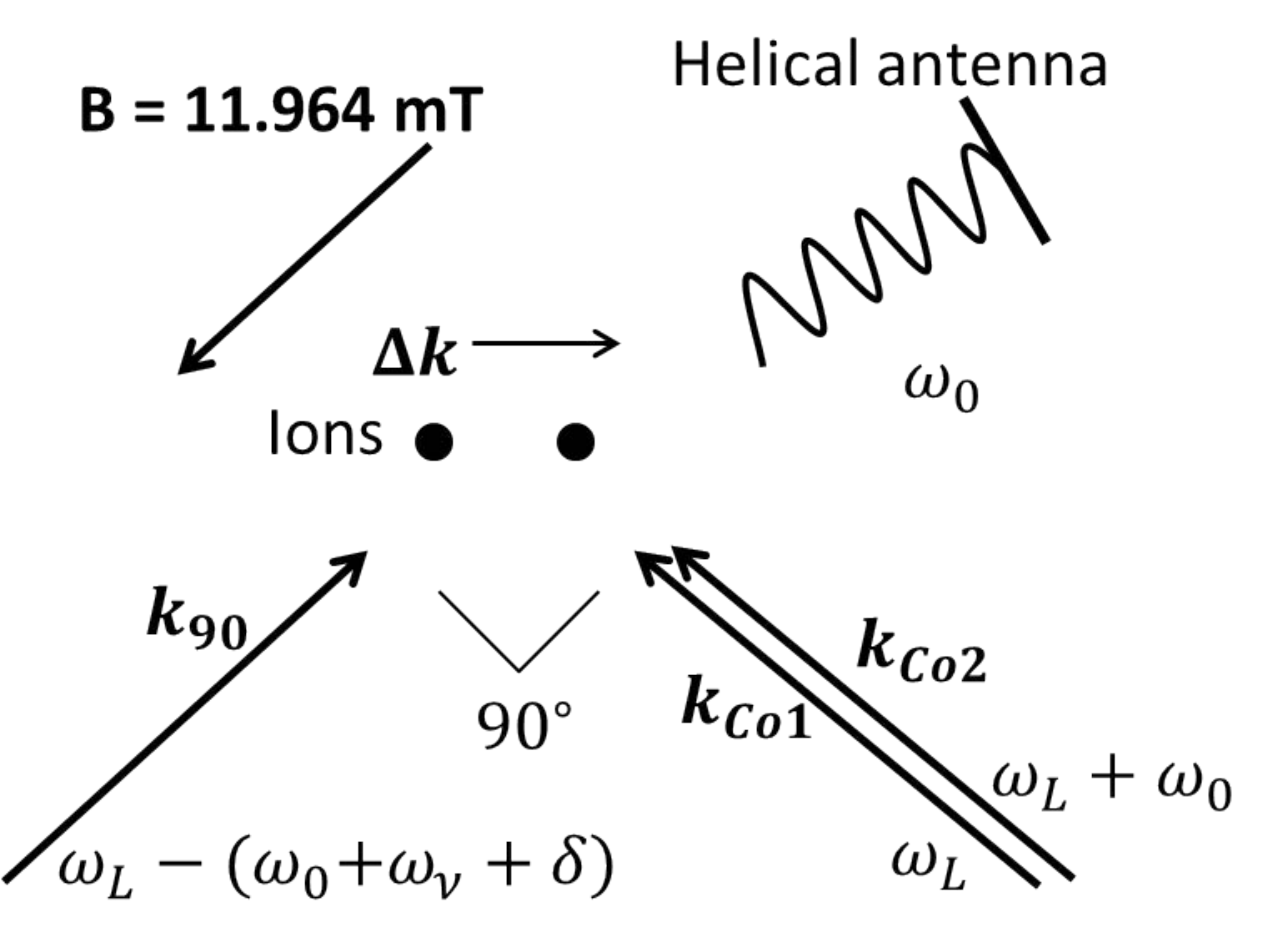}
\caption{ Experimental setup for the entangling gate. Beams with wave vectors $\bf{k_{Co1}}$ and $\bf{k_{90}}$, with a frequency difference of $\omega_0+\omega_\nu+\delta$, are used to non-resonantly drive a sideband transition. Their alignment provides a wave-vector difference $\bf{\Delta k_z}$ such that only vibrational modes along the trap axis are excited. Carrier transitions can be driven either with co-propagating laser beams having wave-vectors $\bf{k_{Co1}}$ and $\bf{k_{Co2}}$, or with a microwave field introduced with a helical antenna.}
\label{BeamDiagram}
\end{figure}

The ions are first Doppler-cooled and optically pumped to the $\left|F=2,m_F=2\right>$ state, followed by Raman sideband cooling of the axial COM and stretch modes to $\bar{n}$ of $\sim0.2$ and $\sim0.05$, respectively. Then, a composite pulse sequence induced by co-propagating beams $\bf{k_{Co1}}$ and $\bf{k_{Co2}}$ (Fig. \ref{BeamDiagram}) is applied, consisting of three resonant pulses $\frac{\pi}{2}_x$, $\pi_y$, and $\frac{\pi}{2}_x$, where the subscript denotes the axis of rotation, transfers each ion from the $\left|2,2\right>$ state directly to the $\left|\downarrow\right>$ state. To perform the gate, we use the Raman laser beam configuration shown in Fig. (\ref{BeamDiagram}). A pair of Raman laser beams, labeled by $\bf{k_{Co1}}$ and $\bf{k_{90}}$ with difference wave vector $\bf{\Delta k_z } = 2\sqrt{2} \pi/\lambda$ along the trap axis is used to drive the detuned sideband transition. In this configuration, only the vibrational modes along the trap axis interact with the laser beams. The beat-note frequency of these two laser beams is blue detuned by a frequency $\delta$ (typically a few kilohertz) from the stretch mode blue sideband transition. The interaction between this pair of laser beams with the carrier and COM mode sideband transitions can be neglected to a high degree as the beat note is sufficiently detuned from the frequencies for these transitions.

\begin{figure}
\includegraphics[width=\linewidth]{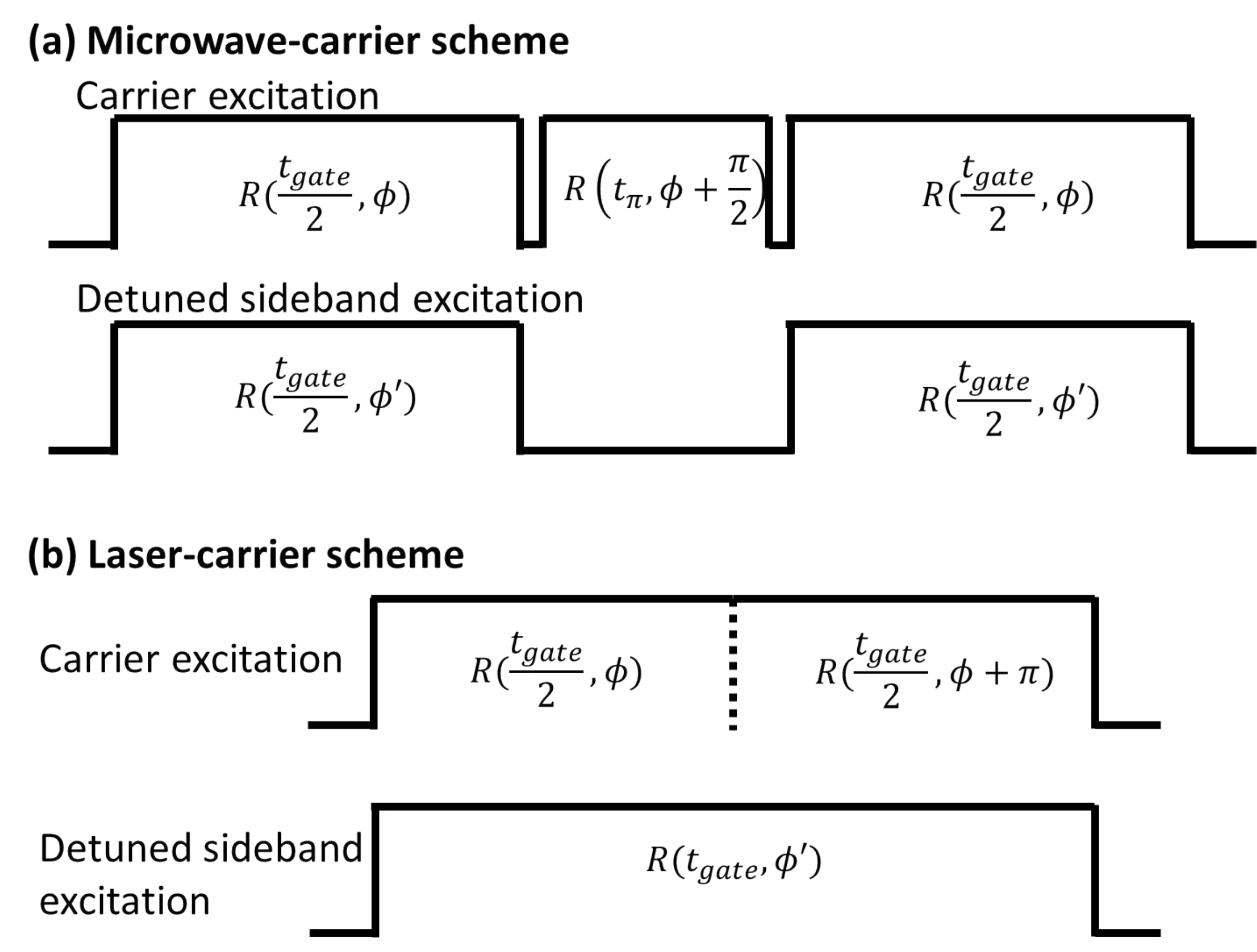}
\caption{Gate timing sequence. For the microwave-induced-carrier gate we perform a $\pi$ rotation with a $\pi/2$-phase shift that refocuses the fast spin oscillations induced by the carrier and suppresses errors in the gate timing \cite{Hayes2012}. For the laser-induced-carrier gate we switch the phase of the carrier by $\pi$ during the second half of the sideband drive.}
\label{ComputationSequence}
\end{figure}

Simultaneous with the detuned sideband excitation, a microwave field with frequency $\omega_0 /2\pi = 1.207$ GHz applied by use of a helical antenna was used to drive carrier transitions, and we achieve carrier $\pi$-pulse durations of approximately 11 $\mu$s \cite{microwavecoupling}. To remove the dependence of the final state on the carrier Rabi frequency, we perform a spin-echo type sequence. In this case, we apply a $\pi$-pulse with a $\frac{\pi}{2}$-phase shift with respect to the carrier in the middle of the gate sequence (Fig. \ref{ComputationSequence}(a)). This pulse has the further benefit of suppressing errors in the gate detuning $\delta$ and gate duration that can lead to residual spin-motion entanglement at the end of gate operation. \cite{Hayes2012}. The total gate interaction duration (not including the additional carrier $\pi$ pulse) is equal to $\frac{4 \pi}{\delta} = 250$ $\mu$s.

\begin{figure}
\includegraphics[width=\linewidth]{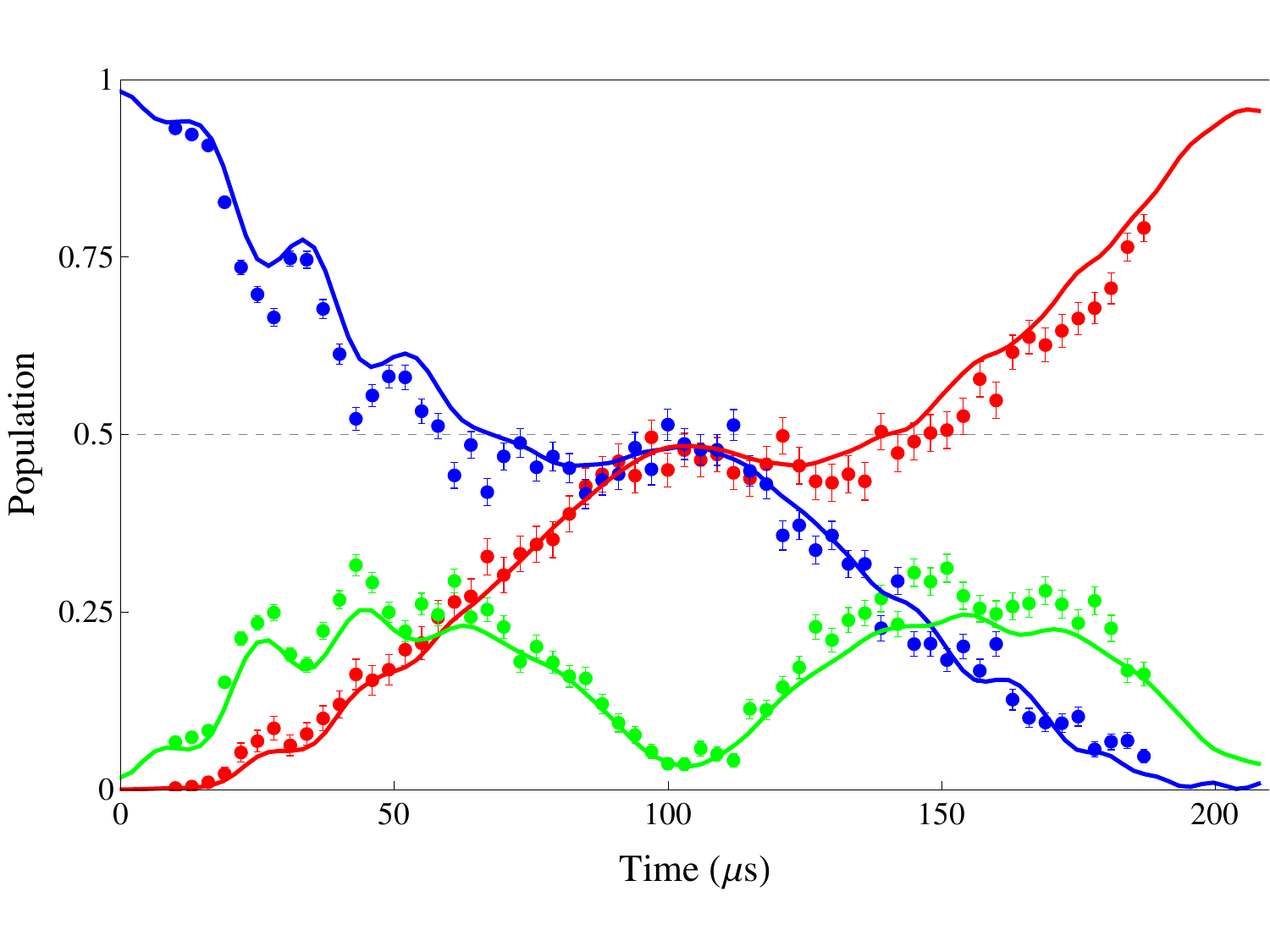}
\caption{Evolution of the populations of $\left|\downarrow\downarrow\right>$ (blue points), $\left|\uparrow\uparrow\right>$ (red) and anti-aligned spin states (green) as a function of the duration of simultaneous application of laser-induced carrier and detuned sideband excitation. The phase of the carrier is shifted by $\pi$ at half of the interrogation time for each point. The gate time for this case is approximately $105$ $\mu$s, at which point the Bell state $\Psi_{Bell}=\frac{1}{\sqrt{2}}\left(\left|\downarrow\downarrow\right>+\left|\uparrow\uparrow\right>\right)$ (in the ideal case) is created. The solid lines show the results of simulation that include a combined spontaneous emission error of 0.019 at the gate time, and state preparation and detection error of 0.017. Error bars are standard errors determined from the measured standard deviation of the points. }
\label{GateScan1}
\end{figure}

We also perform the gate by use of a laser-induced carrier excitation. The frequency of beam $\bf{k_{Co2}}$ is adjusted to drive the carrier excitation together with $\bf{k_{Co1}}$ through a stimulated-Raman process. The carrier $\pi$-transition induced by these two beams has a duration of approximately $5$ $\mu$s. As these two beams are co-propagating, the Rabi frequency is highly immune to ion motion. In this case, the carrier drive was continuously applied with a spin-echo $\pi$-phase shift applied halfway through the gate. In contrast to the microwave case, the $\pi$-phase shift corrects only for errors in the carrier Rabi frequency but does not suppress errors that lead to residual spin-motion entanglement. As a result, the laser-induced-carrier gate could be accomplished in a shorter duration of $\frac{2\pi}{\delta} = 105$ $\mu$s. Fig. \ref{GateScan1} shows the populations evolution induced by the gate Hamiltonian as a function of laser interrogation duration. All laser beams are generated from a single laser source with a wavelength near 313 nm, red detuned $\sim$ 260 GHz ($\sim$ 160 GHz) from the $^2S_\frac{1}{2}$ to $^2P_\frac{1}{2}$ transition for the microwave (laser)-induced-carrier gate.

Read-out of the ion states in the $\left|\downarrow\right>$, $\left|\uparrow\right>$ basis at the end of gate operation is accomplished with state-dependent resonance fluorescence. First, the $\left|\downarrow\right>$ state is transferred to the $\left|2,2\right>$ state by use of the same composite pulse sequence as for the state initialization, and the $\left|\uparrow\right>$ state is transferred to the $\left|1,-1\right>$ state by use of a single $\pi$-pulse. A $\sigma^+$ polarized beam tuned to the $^2S_{\frac{1}{2}}\left|2,2\right>$ $\leftrightarrow$ $^2P_{\frac{3}{2}}\left|3,3\right>$ is then applied for 250 $\mu$s. Ions excited to the $^2P_{\frac{3}{2}}\left|3,3\right>$ state can only decay back to the $^2S_{\frac{1}{2}}\left|2,2\right>$ state so this transition is closed and the ions cycle between these two states \cite{JostThesis}. A fraction of the emitted photons from this transition are collected and register an average of $\sim30$ photons per ion in a photomultiplier tube. Ions in the $\left|1,-1\right>$ state scatter almost no photons (three average background counts are registered during the detection period due to stray scattered light). Detection counts yield three possible outcomes: two ions bright ($\left|\downarrow\downarrow\right>$), one ion bright ($\left|\uparrow\downarrow\right>$ and $\left|\downarrow\uparrow\right>$) , and zero ions bright ($\left|\uparrow\uparrow\right>$). The probabilities of those outcomes, $P_2$, $P_1$, and $P_0$ respectively, are determined by fitting a triple Poissonian function to the histogram of counts obtained in each experiment.

\begin{figure}
\includegraphics[width=\linewidth]{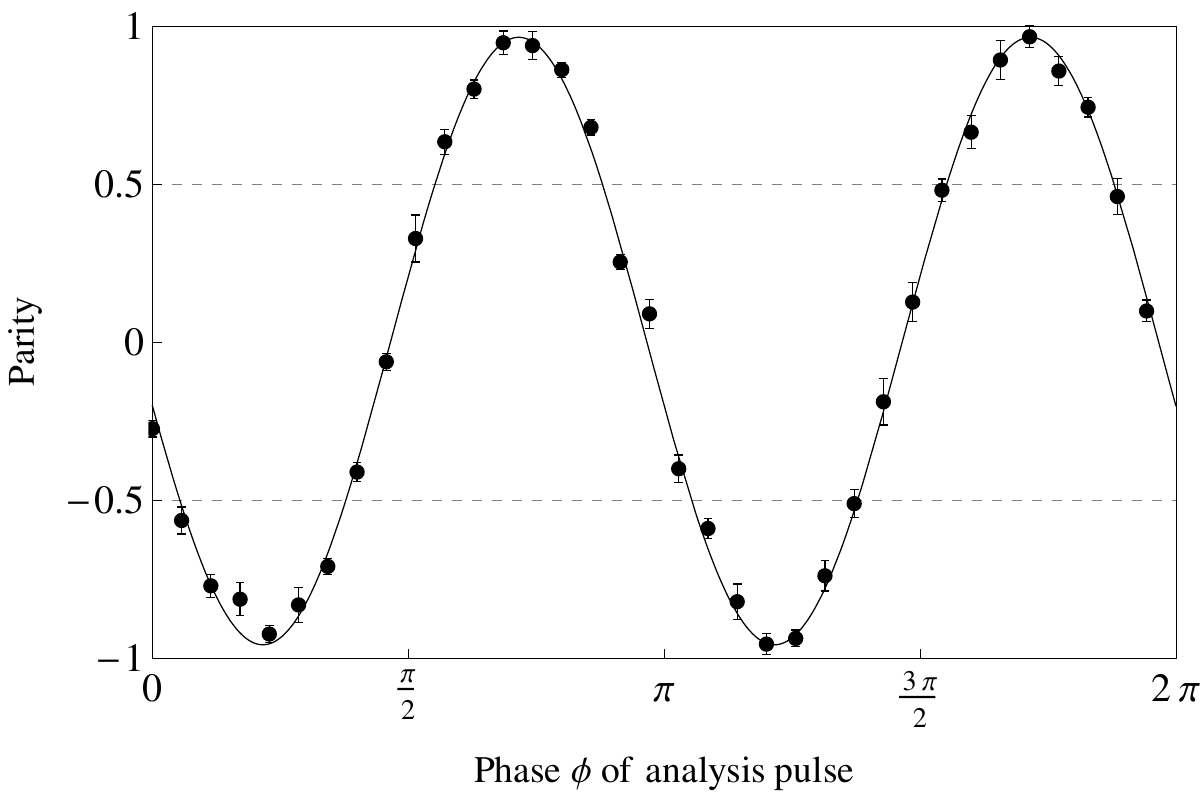}
\caption{The parity oscillation of the microwave-induced-carrier gate is obtained by adding an analysis $\pi/2$-pulse with a variable phase, $\phi$, at the end of the gate and measuring the parity, $P_2+P_0-P_1$, of the qubit populations. The contrast of the parity oscillation is extracted to be $0.960(8)$ by fitting $A\cos(2\phi+\phi_0)+B$ to the data points. Together with a population measurement of $P_2+P_0 = 0.988(4)$ after the gate, a fidelity of $0.974(4)$ is deduced. Error bars are standard errors determined from the error in the fits to the fluorescence count histograms. }
\label{Parity1}
\end{figure}

Each experiment begins with the ions in the $\left|\downarrow\downarrow\right>$ state, with the gate ideally creating $\Psi_{Bell}=\frac{1}{\sqrt{2}}\left(\left|\downarrow\downarrow\right>+\left|\uparrow\uparrow\right>\right)$. The performance of the gate is characterized by measuring the state fidelity, which is given by $\langle \Psi_{Bell}|\rho_{exp}|\Psi_{Bell}\rangle$ where the density matrix $\rho_{exp}$ describes the experimentally produced state. The $\downarrow\downarrow$ and $\uparrow\uparrow$ diagonal elements of the density matrix are determined from $P_2$ and $P_0$. The off-diagonal elements $\rho_{\downarrow\downarrow,\uparrow\uparrow}$ are determined by applying an analysis carrier $\pi/2$-pulse with a variable phase $\phi$ to the ions and fitting the resulting oscillation of the parity ($P_2+P_0-P_1$) to the function $A \mathrm{cos}(2\phi+\phi_0)$ (Fig. \ref{Parity1}). The entanglement fidelity, $F$, is $(P_0+P_2+A)/2$ \cite{Sackett2000}. For the microwave-induced-carrier gate we find $P_0+P_2 = 0.988(4)$ and $A=0.960(8)$, which gives $F = 0.974(4)$. For the laser-induced-carrier gate we find $P_0+P_2=0.961(1)$ and $A = 0.930(8)$, which gives $F = 0.946(4)$.


The main sources of infidelity for creating the Bell state are enumerated in Table \ref{ErrorBudget}. The Raman laser beams off-resonantly excite the $^2S_{\frac{1}{2}}$ to $^2P_{\frac{1}{2}}$ and $^2P_{\frac{3}{2}}$ transitions and scatter photons incoherently. Errors caused by spontaneous emission \cite{Ozeri2007} (determined by applying the Raman beams sufficiently detuned from the carrier or sideband transitions) are given in line 1 of Table \ref{ErrorBudget} for both the microwave-induced-carrier and laser-induced-carrier gates. In the latter case, spontaneous emission from the beams used to drive the carrier transition dominates the scattering error. For the microwave-induced-carrier gate, the scattering error is further reduced by increasing the detuning of the Raman lasers by an additional $\sim$ 100 GHz.

The combined errors for state preparation and detection, including transferring into and out of the qubit manifold are given in line 2 of Table \ref{ErrorBudget}. The improved state preparation and detection for the microwave-induced-carrier gate was achieved by more careful calibration of the laser beams' polarization and alignment.

Errors due to fluctuations of the carrier Rabi frequency that are slow compared to the gate duration are suppressed by spin-echo techniques. However, fluctuations on the time scale of the gate duration cause error. We can approximately characterize this error, given in line 3 of Table \ref{ErrorBudget}, by performing the gate sequence with only the carrier drive applied and measuring the probability to end in the $\left|\downarrow\downarrow\right>$ state for the laser-induced carrier or the $\left|\uparrow\uparrow\right>$ state for the microwave-induced carrier. We estimate gate error caused by fluctuations in the laser-induced sideband excitation due to laser intensity and beam-pointing fluctuation to be  approximately $10^{-3}$. This error is determined by performing a Monte-Carlo simulation incorporating measured laser-intensity and beam-pointing fluctuations of the sideband excitation.

Estimated errors caused by motional heating from electric field noise \cite{Sorenson2000}, and errors caused by fluctuating Debye-Waller factor associated with the COM mode due to its finite Lamb-Dicke parameter and finite thermal energy \cite{bible,Sorenson2000} are given in line 4 of Table \ref{ErrorBudget}. As the gate duration for the microwave-carrier-gate was longer, it suffered more from motional heating effects, resulting in a larger error compared to the laser-carrier-gate. The finite thermal energy and motional heating of the stretch mode lead to a much smaller error of $<10^{-3}$.

A further source of error is due to the fast oscillations caused by the second term in Eq. (\ref{dressedStatePic}), which are neglected in the rotating-wave approximation, with values given in line 5 of Table \ref{ErrorBudget}. This error is determined by performing numerical simulations with experimental values as input parameters, assuming they are noise free. This error is reduced for larger $\Omega_{C}$ and an error of $10^{-4}$ can be achieved for $\Omega_{C}=$ $\sim40$ $\delta$ for the case where the gate duration is $\frac{2 \pi}{\delta}$.

\begin{table}
\begin{tabular}{|l|c|c|c|}
  \hline
  & Infidelities/Errors & Microwave & Laser \\
  \hline
  1. & Spontaneous emission& $2.8\times 10^{-3}$ & $19\times 10^{-3}$ \\
  2.  & State preparation and detection & $9.1\times 10^{-3}$ & $17\times 10^{-3}$ \\
  3. & Carrier drive infidelities& $1.3\times 10^{-3}$ & $16\times 10^{-3}$ \\
  4. & Heating and motion fluctuation& $\sim 10\times 10^{-3}$ & $\sim 6\times 10^{-3}$ \\
  5. & Fast oscillation term& $\sim 3 \times10^{-3}$ & $<10^{-3}$ \\
  6. & Imperfect sideband drive& $\sim 10^{-3}$ & $\sim 10^{-3}$ \\
  \hline
\end{tabular}
\caption{Error budget of the gate for microwave-induced and laser-induced carrier excitation.}
\label{ErrorBudget}
\end{table}

With better microwave delivery \cite{Ospelkaus2008,Ospelkaus2011}, the carrier Rabi frequency could be significantly increased and stabilized, reducing errors from the second term of Eq. (\ref{dressedStatePic}). Larger carrier Rabi frequency also allows the gate to be performed faster, reducing the effect of motional heating on the gate performance. The errors due to motional heating can be further reduced by increasing the normal mode frequency, and by cleaning the ion trap electrodes to reduce the electric-field noise responsible for motional heating \cite{Hite2012}. The spontaneous emission error can be reduced to the $10^{-4}$ level by further increasing the Raman laser detuning \cite{Ozeri2007}. Randomized benchmarking could be used to better evaluate the performance of the gate without state preparation and detection error. \cite{Knill2008,Gaebler2012}. With the above improvements the gate errors could be reduced below $10^{-4}$ \cite{Lemmer2013}.

\section{Acknowledgements}

\begin{acknowledgments}
We thank A. Wilson and D. Slichter for comments on the manuscript. J. P. G. acknowledges support by NIST through an NRC fellowship. This work was supported by IARPA, under ARO contract no. EAO139840, ONR and the NIST Quantum Information Program. This paper is a contribution by NIST and not subject to U.S. copyright.
\end{acknowledgments}


\begin{thebibliography}{10}
\bibitem{microwavecoupling}
The antenna was impedance matched to the source at the qubit frequency. Other frequencies have poor coupling to the ions and thus microwave transition was not used for mapping in and out of the qubit manifold.
\end{thebibliography}
\end{document}